\documentclass[11pt,twoside]{article}
\usepackage{asp2010}

\resetcounters

\def\msun{\hbox{M$_\odot$}}
\def\sigma_sfr{\hbox{$\Sigma_{\rm ~SFR}$}}
\def\mvb{\hbox{M$_{\rm V}^{\rm brightest}$}}

\newcommand{\tlu}{T\ensuremath{_{L}}(U)\ }

\begin{document}

\title{Young Massive Clusters and Their Relation to Star Formation}
\author{Nate Bastian
\affil{$^1$Excellence Cluster Universe, Boltzmannstr. 2, 85748 Garching, Germany}}

\begin{abstract}

The formation of massive stellar clusters is intricately linked to star formation on local and global scales.  All actively star forming galaxies are forming clusters, and the local initial conditions likely determining whether bound massive clusters or unbound associations are formed.  Here, we focus on observed scaling relations between cluster populations and the properties of the host galaxy.  In particular, we discuss the relations between the fraction of U-band light from clusters vs. their host galaxy as well as the brightest cluster vs. population size and host galaxy star formation rate (SFR).  We also discuss the the fraction of stellar mass formed within bound clusters within the Galaxy, nearby dwarf galaxies, as well as starbursts and mergers.  Bound clusters appear to represent $\sim10$\% of star formation within most galaxies, although there are intriguing hints that this fraction systematically increases in galaxies with higher star formation rate surface densities.  Throughout the review we highlight potential avenues for future study.

\end{abstract}

\section{Introduction}

Since the advent of the {\em Hubble Space Telescope} (HST) it has been clear that all intensely star forming galaxies host stellar clusters with masses comparable to those of the ubiquitous globular clusters (e.g., \citealt{holtzman92}).  However, their blue colours imply ages less than $\sim$1~Gyr suggesting that their formation is directly linked with the intense star forming events that their hosts are undergoing.  The post-starburst/merger galaxy NGC~7252 is often taken as a template, as it hosts an enormous cluster population that has an age distribution that peaks during the inferred merger between the host spirals (e.g., \citealt{miller97,schweizer98}).  

In this review, our aim is to link the properties of the cluster population with that of the host galaxy, with a focus on young massive clusters (YMCs), i.e. clusters with masses above $\sim10^4$\msun\ and ages less than $\sim1$~Gyr (see \citet{pz10} for a recent review), although we note that the continuous age and mass distributions of clusters makes this distinction somewhat arbitrary.  In order to begin we need a working definition of a cluster.  For this we will adopt the definition put forward by \citet{gieles11}, that a cluster is a stellar grouping whose age is greater than a current dynamical crossing time.  This naturally includes gravitationally bound stellar groupings and excludes unbound expanding associations.  One implication of this definition is that young objects (commonly referred to as embedded clusters) are not included as they are not yet old enough to have had their stars cross from one side to the other.  This avoids the complication of defining ``clusters" at young ages, where we see a continuous surface density distribution \citep{bressert10}, i.e. where there is not a clear separation between ``clustered" and ``distributed" stars \footnote{\citet{schweizer06} summarised this confusion well.  He writes:  "In my opinion, it is unfortunate that this loose, non-astronomical use of the word ÒclusterÓ may reinforce an increasingly popular view that most stars form in clusters. By the traditional astronomical definition of star clusters as gravitationally bound aggregates, most of the objects tallied ... in The Antennae are not clusters, but likely young stellar associations. It seems to me in much better accord with a rich body of astronomical evidence gathered during the past 50 years to state that--although {\em star formation is clearly clustered}--even in mergers gravitationally bound clusters (open and globular) form relatively rarely and {\em contain $<10$\% of all newly-formed stars}."}.

Recent dynamical studies of resolved young massive clusters in the Galaxy (e.g., \citealt{cottaar12}) and in the LMC \citep{vincent12} have shown that they are in dynamical equilibrium from a very young age, suggesting that once clusters emerge from their embedded state they are long lived objects.  However, in some cases further cluster disruption must be taken into account when studying the properties of cluster populations (e.g., \citealt{bastian11}) and we will explicitly mention when such corrections have been applied.  

The goal of the studies discussed in this review, is to compare cluster (population) properties with those of the host galaxies.  However, for many of the interesting galaxies (e.g., high SFR galaxies) which are relatively distant (tens of Mpc), it is difficult to distinguish between bound clusters and unbound associations.  After 10-20~Myr, it is relatively easy to make the distinction between between bound and unbound stellar groupings (\citealt{gieles11, bastian12}), but since young stellar clusters/associations outshine their older cousins (hence will be over-represented in any sample), unbound associations may contaminate many of the samples discussed here.  This caveat will be discussed throughout.

In the following sections we will discuss three empirical relations that have been found between individual YMCs, full populations, and their host galaxies.  In \S~\ref{sec:tlu} we discuss the fraction of U-band light coming from clusters relative to the host galaxy's SFR surface density (\sigma_sfr), while in \S~\ref{sec:brightest} we focus on the relation between the brightest cluster within a population and the galactic SFR.  Finally, in \S~\ref{sec:gamma} we review attempts to derive the fraction of star formation that happens in bound clusters as a function of galactic properties ($\Gamma$ - \citealt{bastian08}).


\section{The Fraction of U-band Light in Clusters}
\label{sec:tlu}

Using the {\em Faint Object Camera} onboard HST, \citet{meurer95} measured the fraction of {\it UV} light coming from compact objects compared to the full galaxy in a sample of starburst galaxies.  The {\it UV} traces the light from young, high mass stars (although there is a significant component from $>100$~Myr old stars) so this should be a proxy for the fraction of stars forming in clusters. Meurer et al. report fractions ranging from $\sim20-50$\%, along with a trend for this fraction to increase with the star formation rate surface density ($\sigma_sfr$).  \citet{zepf99} and \cite{adamo11a} found that the ongoing galaxy merger NGC~3256 and blue compact galaxies (BCGs), respectively, also match the Meurer et al. relation.  At face value these results suggest that a higher fraction of stars form in clusters in galaxies with high \sigma_sfr.

There are two important caveats to these results.  The first is that the studied galaxies span a large range of distances from $\sim4$ to tens of Mpc. In nearby galaxies (with generally low \sigma_sfr) individual clusters can be resolved, while for the distant starburst galaxies we are only be able to resolve cluster complexes with potentially a large amount of inter-cluster light.   \citet{adamo11a} discuss this caveat in more detail, although a detailed resolution study has not yet been undertaken.  The second caveat to this result is that it assumes that the average extinction towards clusters is the same as it is to the field, which given the large amounts of wind blown bubbles around massive clusters, may not be justified.

\begin{figure}[!ht]
\plotone{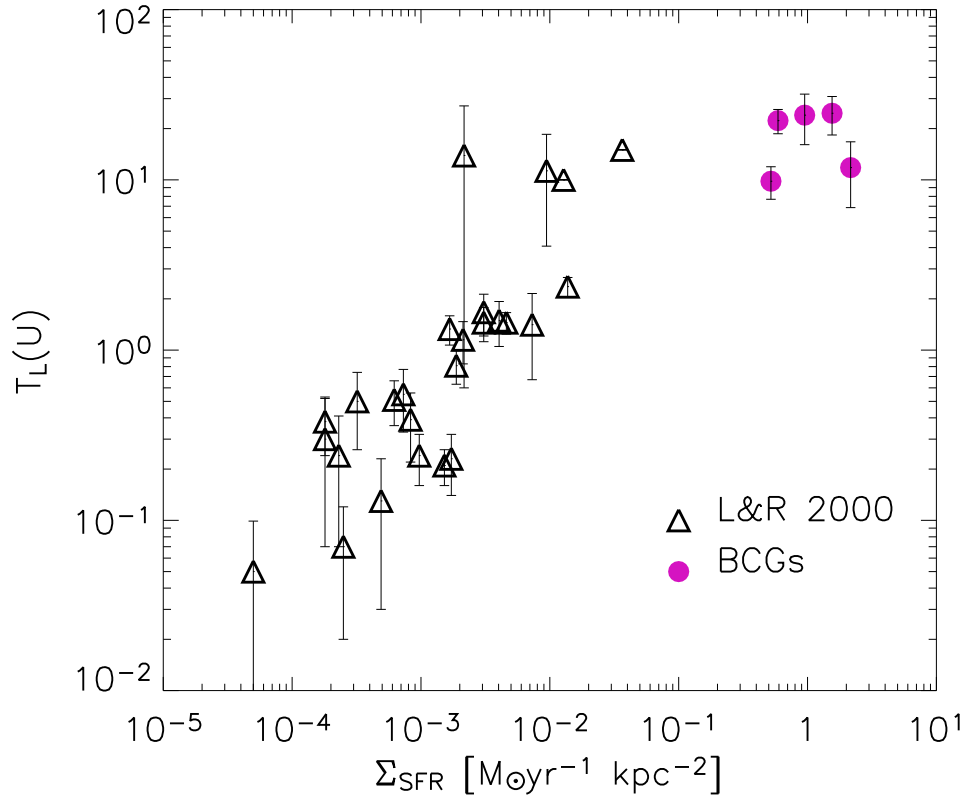}
\caption{{The relation between the percentage of U-band light coming from clusters compared the full galaxy as a function of \sigma_sfr.  The figure is taken from \citet{adamo11a} and includes the data from \citet{larsen99c}.
}}
\label{fig:tlu}
\end{figure}

\citet{larsen99c} extended the Meurer et al. study by looking at a sample of nearby dwarf and spiral galaxies.  They used ground based multi-band data to select young (few hundred Myr or less, but avoiding H${\alpha}$ emitting sources) clusters over the full extent of the galaxies.  They defined \tlu to be the percentage of light in the U-band coming from the clusters in their sample compared to the full galaxy, i.e. $T_L = 100 * L_{\rm clusters}/L_{\rm galaxy}$, known as the specific luminosity.  Their data is shown in Fig.~\ref{fig:tlu} as open triangles, and BCGs from Adamo et al. are shown as filled circles.  Both samples are consistent with the trend observed by \citet{meurer95} for starburst galaxies.  Because the galaxies in the Larsen \& Ritchler sample are all relatively nearby, they should not suffer from the first caveat discussed above, however the second caveat remains.

While the \tlu is a good proxy for the fraction of star formation happening in bound clusters, due to it being a purely observational quantity with limited assumptions, other methods have been developed, which are discussed in turn below.  No theoretical study has investigated the \tlu\ vs. \sigma_sfr\ relation in detail.  Hence it is not clear to what extent \tlu\ traces the fraction of star formation happening in bound clusters or how much the star formation history of the galaxy, cluster disruption, or differential extinction affect the observed values.

\section{The Brightest Clusters vs. Star Formation Rate}
\label{sec:brightest}

Many of the properties of young cluster systems are dominated by size-of-sample effects \citep{whitmore03,larsen04}, meaning that the larger the cluster population is, the more likely it is to sample the extreme ends of the distributions (i.e. forming very bright, high mass, or large radii clusters).  This can be seen in the relation between the luminosity of the brightest cluster (\mvb) vs. the number of clusters (above a certain magnitude cut) within the population (e.g., \citealt{whitmore03}).

\begin{figure}[!ht]
\plotone{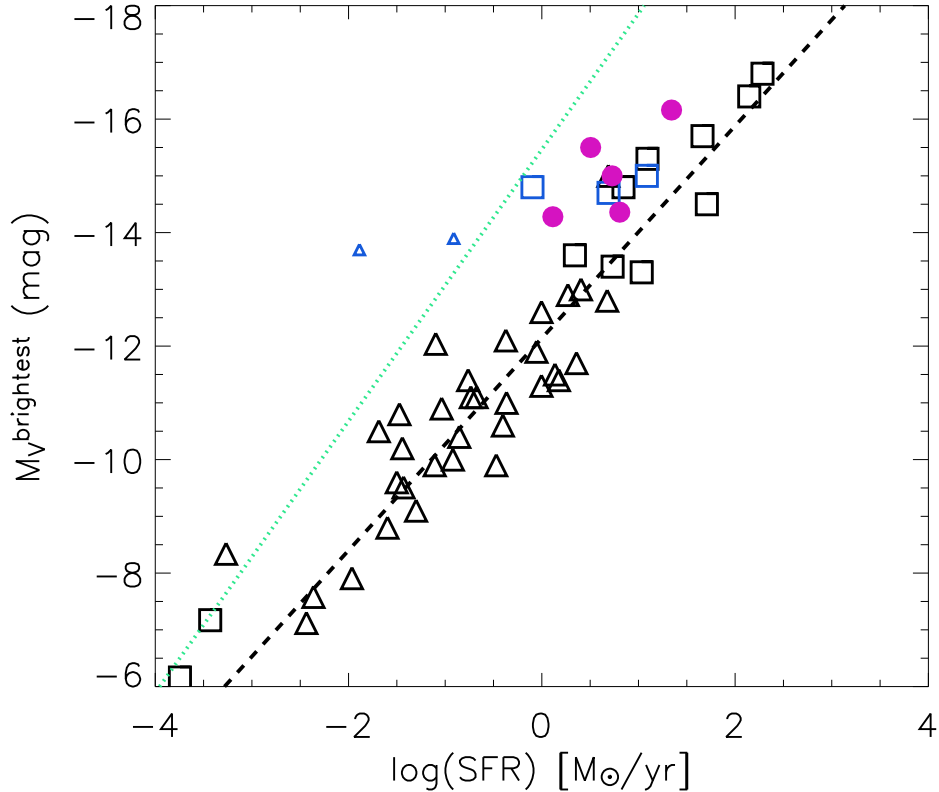}
\caption{{The SFR vs. \mvb\ relation for a sample of galaxies.  The plot is reproduced from \citet{adamo11a}.  The dashed line is the best fit through the \citet{larsen02} data (open triangles) while the (green) dotted line represents the expectations from a series of monte carlo models with all star formation happening in clusters and a power-law mass function with index -2 \citep{bastian08}.
}}
\label{fig:mvb}
\end{figure}

\citet{larsen02} found a relatively tight relation between \mvb\ and the star-formation rate (SFR) of the host galaxy.  His data are shown as open triangles in Fig.~\ref{fig:mvb}, and the dashed line is the best linear fit through the data \citep{weidner04}.  \citet{bastian08} expanded the relation by adding clusters/galaxies at the extreme ends of the distributions (open squares), and \citet{adamo11a} placed their BCGs on the diagram as well (filled circles).  As can been seen there is a clear relation that holds for over six orders of magnitude in SFR, with high SFR producing brighter most luminous clusters.

A nice feature of size-of-sample effects is that it is straightforward to model them using monte carlo techniques.  \citet{bastian08} carried out such a study by generating large numbers of cluster populations.  This was done as follows.  First, assume that all stars form in clusters and that the clusters are sampled from a mass distribution, either a pure power-law with index, -2, or with a Schechter function (a power-law at the low mass end, followed by an exponential distribution above a certain mass).  Then, randomly assign each cluster an age between 0 and 100~Myr in order to mimic a constant cluster formation history.  The SFR of each population can be found by dividing the total mass formed in clusters by the duration of the experiment (100~Myr).  Each cluster in the population was then assigned a luminosity and colour, by adopting simple stellar population (SSP) models of the appropriate age and mass.  The most luminous cluster was then chosen in each population, and plotted against the SFR for that model.

The mean result for a pure power-law mass function (index $-2$) is shown in Fig.~\ref{fig:mvb} as a (green) dotted line.  The slope of the line is controlled by the form of the underlying cluster mass function, with lower power-law indices (i.e., indices below $-2$) resulting in shallower slopes.  However, one generic feature of all the simulations, regardless of the form of the underlying mass function, was that they always lay above and to the left of the observations.  The simplest interpretation of this is that not all star formation is happening in clusters.  We note that the scatter observed in Fig.~\ref{fig:mvb} is consistent with that expected from stochastic sampling of the cluster mass function.

In order to shift the models onto the observations, \citet{bastian08} defined $\Gamma$ to be the fraction of star formation that happens within bound clusters.  They assumed that due to the relatively young ages of the clusters used in Fig.~\ref{fig:mvb} (see \citet{larsen09} and \citet{gieles09} for a more in depth discussion of the age and brightness of the most luminous cluster of a population) and the inferred high masses of the clusters, that cluster disruption was not significantly affecting the observed relation.  With this assumption, they found that $\Gamma \sim 0.08$, with no strong dependence with the host galaxy SFR.

\citet{adamo11a} note that their BCG galaxies lie systematically above the best fit to the observations, suggesting systematically higher $\Gamma$ values.  Additionally, \citet{cook12} found that within their sample of 37 dwarf galaxies, those galaxies that had clusters (7) were consistent with the trend observed in Fig.~\ref{fig:mvb}.  However, a handful of their galaxies were expected to host identifiable clusters based on their observed SFR, although none were found, which was surprising even taking stochastic sampling of the cluster mass function into account.  If true, this would point to an additional environmental effect which either allows or stops clusters from forming.

Like in the previous relation, resolution effects may influence the observed trend.  Most of the galaxies shown in Fig.~\ref{fig:mvb} were studied with HST imaging, or were nearby enough to (semi)resolve clusters with ground based imaging.  However, if more distant systems are included with ground based (lower resolution) imaging, they will appear systematically brighter (since cluster formation is in itself a clustered process, hence full complexes of clusters can be mistaken for a single cluster).  Hence care must be taken before adding new sources to the diagram.

A possible avenue for future work, beyond adding new galaxies, is to extend the relation into different bands (i.e. the near-IR) to limit the effects of extinction, although the presence or absence of red super giants within the clusters could complicate the analysis.  Additionally, including not just the brightest cluster but the median (or total) luminosity of the top 5 or 10 brightest clusters could significantly reduce the scatter in the observed relation and hence allow one to look at more subtle variations of $\Gamma$ with host galaxy SFR.

\section{The Fraction of Star Formation In Clusters}
\label{sec:gamma}

While the above relations between \tlu,  \mvb, and the properties of the host galaxy provide some insight into the cluster formation process, their are only proxies for $\Gamma$ (the fraction of star formation that happens within bound clusters).  A number of studies have attempted to estimate $\Gamma$ within different environments, and while a number of caveats remain, the results appear to be converging to a consistent picture.

There are a number of ways to estimate $\Gamma$, each with their own caveats, but the basic process is to use the observed portion of a cluster population (old and/or low mass clusters are faint and hence will not be observable, even if they are present in large numbers) to estimate the cluster formation rate (CFR) and compare that to the current star formation rate (SFR) through H$\alpha$ or IR observations or through the reconstructed star formation history made through resolved stellar populations.

\subsection{Galactic studies}

The most accessible region to estimate this fraction is in the solar neighbourhood. These studies are based on open cluster catalogues, along with estimates of the total local star-formation rate. While we can study nearby clusters in extreme detail, current cluster catalogues are only complete to $\sim800$~pc \citep{piskunov08}, and were significantly more incomplete in previous decades. 

As noted in the introduction, we will not discuss the fraction of stars forming in ``embedded clusters", as this depends largely on the definition adopted to separate ``clusters" from ``distributed" stars \citep{bressert10}.

Most early studies (e.g., \citealt{roberts57}) focussed on the fraction of O and B stars that form within clusters {\em or} associations relative to a purely isolated environment.  They found that the vast majority are formed in groups.  \citet{von68} estimated the cluster and star formation rate in the solar neighbourhood based on then current cluster catalogues and concluded that $\sim20$\% of star formation happened in bound clusters, while \citet{wielen71} suggested that only a relatively small fraction (of order $2-5$\%) of stars form in open clusters.  Most subsequent studies, which are fundamentally based on the same technique of estimating the CFR and SFR, have converged on a value of $\sim10$\% \citep{miller78,adams01,lamers06}.

One outlying study was that of \citet{piskunov08} who found a much higher value of $\Gamma \sim40$\%.  This was based on a more complete catalogue of clusters.  However, one essential ingredient in all such studies is the accurate estimation of each cluster's mass and age.  These authors did not use the usual method of counting the number of stars within a cluster (above some magnitude cut) or comparing the cluster absolute magnitude to SSP models (of the appropriate age) to estimate each clusters mass.  Instead, they fit King profiles to confirmed cluster members and estimated the mass based on the truncation radius of the best fit profile.  This method is extremely sensitive to any error in the truncation radius (to the third power) and it also assumes that young clusters are tidally truncated.  Comparing their derived masses for a number of well studied clusters (e.g., the ONC) with in depth photometric studies suggests systematic deviations, which could explain the high value of $\Gamma$.  Using the Piskunov et al. sample, but with masses estimated on the number of stars (and age) above a certain magnitude limit for each cluster, \citet{lamers06} found that $\sim7$\% of stars formed in bound clusters, in good agreement with the previous studies.


\subsection{Extragalactic determinations}

Outside the Galaxy, relatively few direct estimates of the fraction of stars born in clusters have been made. Using size-of-sample effects (the most massive cluster in different age bins), \citet{gieles08} estimated that optically selected clusters (older than $\sim7$~Myr) in the SMC represented $2 - 4$\% of the total star-formation in the galaxy. By comparing the location of clusters/associations with the location of H$\alpha$ emission, \citet{fall05} report that the majority ($>80$\%) of star formation within the Antennae galaxies happens within clusters {\em or} associations.  However, \citet{schweizer06} suggests that even within galaxy major mergers,  $< 10$\% of star formation bound star clusters.

\begin{figure}[!ht]
\plotone{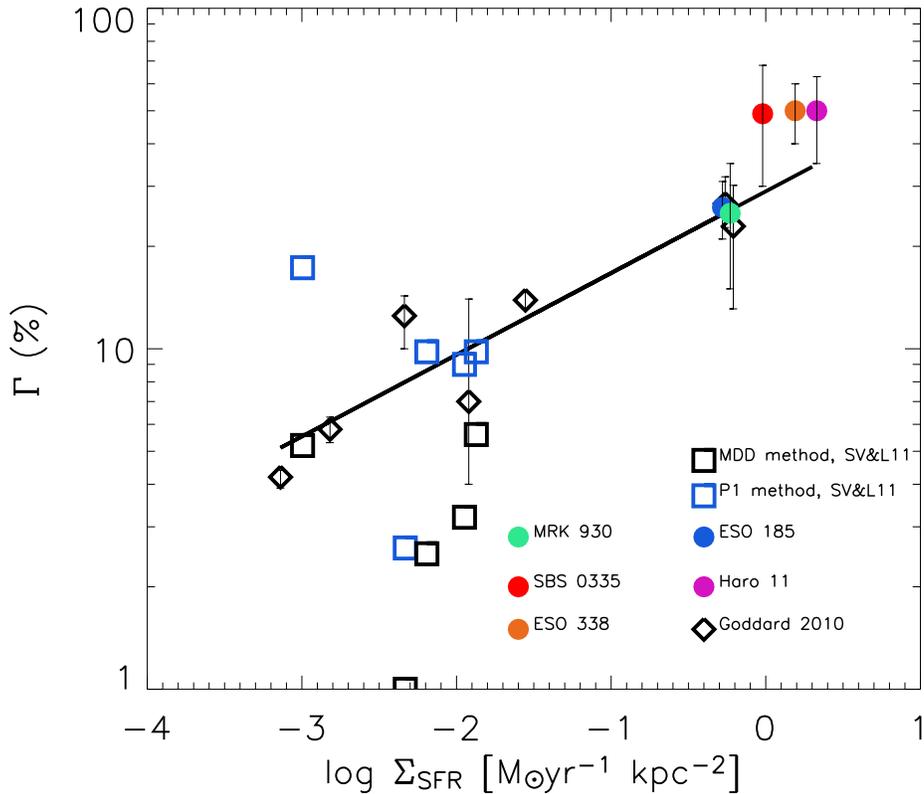}
\caption{{The relation between \sigma_sfr and the fraction of star formation that happens in bound clusters ($\Gamma$).  The solid line is the best fit through the \citet{goddard10} relation.  The plot is reproduced from \citet{adamo11a}.
}}
\label{fig:gamma}
\end{figure}

\citet{goddard10} have estimated $\Gamma$ for a sample of galaxies (SMC, LMC, NGC 1569, M83, and NGC 3256) by extrapolating the observed part of the cluster distribution (i.e., the age and mass of the young and massive clusters) to estimate the total population (they also provide a detailed analysis of the associated errors in converting the observed to the total population), and find an average of $\sim10$\%.  However, they report a trend of increasing $\Gamma$ as a function of increasing \sigma_sfr.  Their data are shown as open diamonds in Fig.~\ref{fig:gamma}, and the reported best fit relation is shown as a solid line ($\Gamma (\%) = 29 \times \Sigma_{\rm SFR}^{0.24}$, where \sigma_sfr is in solar masses per year per kpc$^{-2}$).  Some of the values were calculated only using clusters with ages less than 10~Myr, whereas others were calculated using clusters with ages between 10 and 100 Myr.  Hence, it is possible that if cluster disruption operates strongly over $\sim10$~Myr time scales, this could influence the results.

\citet{esteban11} studied nine regions in a sample of four nearby star-forming spiral galaxies using multi-wavelength HST imaging.  The found values of $\Gamma$ between 2\% and 10\%, in good agreement with previous studies.  Depending on how they corrected for cluster disruption, their values for individual galaxies varied by a factor of 2.  These authors did not find any strong correlation with \sigma_sfr\ (their data are shown open squares, with the two colours representing different disruption corrections), although their galaxies only covered a relatively narrow range of \sigma_sfr.  However, the authors did find a correlation with \tlu, suggesting that \tlu may act as a good proxy for $\Gamma$, although an exact mapping between the two has yet to be reported.  Increasing the sample of galaxies that have had $\Gamma$ estimated and \tlu\ measured could provide a large step forward in our understanding of cluster formation and destruction.

Also shown in Fig.~\ref{fig:gamma} are the BCGs from \citet{adamo11a} (filled circles).  These studies are based largely on young clusters, so that cluster disruption should not be a strong factor.  However, as discussed in \citet{adamo11a}, many of these galaxies are quite distant, meaning that individual clusters cannot be resolved, hence the sources within their catalogues may be cluster complexes.  This would lead to systematically higher $\Gamma$ values.

\citet{cook12} estimated $\Gamma$ for a large number of low SFR dwarf galaxies as part of the ANGST HST survey.  They found that summing their entire sample together led to a $\Gamma$ value of 2--8\%, consistent with the \citet{goddard10} relation.  However, a handful of their sample were inconsistent with expectations, even accounting for stochastic sampling of the cluster mass function, potentially suggesting that a second parameter was relevant in whether a cluster was formed or not.

As noted by the authors, the above studies have a number of caveats associated with them.  There is scope to improve upon this first generation of extragalactic studies, starting with understanding resolution affects and identifying clusters, to the derivations of cluster properties and the assumption of a constant star formation rate of the host galaxy over $\sim100$~Myr timescales.  One such study that is likely to overcome many of these limitations is the The Panchromatic Hubble Andromeda Treasury (PHAT) survey \citep{dalcanton12}.  This survey will cover $\sim1/3$ of M31 with multiple filters and allow the identification of thousands of stellar clusters along with providing estimates of their ages/masses \citep{johnson12}.  The survey will also provide spatially resolved star formation history based on individual resolved stars.  Combining the cluster catalog with the star formation history will allow the determination of $\Gamma$ for different regions of the galaxy to test for any dependence on SFR, \sigma_sfr, galactocentric distance, ISM propertes, etc.

There has been recent theoretical advances in our understanding of $\Gamma$ and its potential variation with environment.  This is beyond the scope of the current review, but we point the interested reader to \citet{kruijssen12} who presents a theoretical framework in order to understand these results, linking the density distribution of the ISM to star/cluster formation.

\section{Concluding Remarks}

While a significant amount of work has been invested into finding and understanding the above relations, the field is still at a relatively youthful stage.  Many caveats remain in the observations/interpretations and theory is just beginning to catch up.

It is interesting to note that some of the observed properties (i.e., \tlu\ and $\Gamma$) correlate with \sigma_sfr\ while others (\mvb\ or the total number of clusters with a population) correlate with SFR.  The reason for this could be that \mvb\ (or total number) is dominated by size-of-sample effects which is more related to the host SFR, i.e., forming more stars results in more clusters which sample the distributions more fully.  For example, doubling the SFR does not change the fraction of stars that end up in clusters, it simply doubles the number of clusters present.  Properties such as \tlu\ or $\Gamma$\ reflect {\em relative} quantities, i.e., what part of the star formation hierarchy ends up in bound clusters, and hence may depend on the physical properties of the host (or more likely the physical properties of the ISM within the host galaxy).  Changing \sigma_sfr\ may influence the ISM, which in turn may affect the distribution of star formation, altering the fraction of star formation occurring in bound clusters.

Finally, we note that the relations discussed here apply only to young clusters within galaxies.  While there are many links between YMCs and globular clusters (c.f., \citealt{pz10}) there are other hints that they may have formed in very different conditions.  One recent example is  a study of the metal poor globular clusters in the Fornax dwarf galaxy.  These clusters represent $\sim25$\% of all the metal poor stars within the galaxy, suggesting an extremely high $\Gamma$ value, a cluster mass function devoid of low mass clusters, and negligible cluster disruption \citep{larsen12}.  Hence, it remains to be seen if and how YMCs can be used as templates for globular cluster formation.

\acknowledgements I would like to thank Diederik Kruijssen and Angela Adamo for insightful discussions as well as the organisers of the conference and staff of the Sterrekundig Instituut Utrecht.

\bibliography{bastian_n}

\end{document}